# A Semi-Empirical Approach to a Physically Based Aging Model for Home Energy Management Systems

C. Miller, M. Goutham, X. Chen, P.D. Hanumalagutti, R. Blaser, S. Stockar

*Abstract*— A growing interest in the study of aging related phenomena in lithium-ion batteries is propelled by the increasing utilization of energy storage systems in electric vehicles and in buildings as stationery energy accumulators paired with renewable energy sources. This paper proposes a mixed-degradation model approach that combines the benefits of a semi-empirical approach with that of a physics-based model. This enables easy calibration for different battery chemistries, the ability to extrapolate when necessary, and is computationally efficient enough to be coupled with real-time running control systems. To demonstrate the effectiveness of the proposed approach, the effect of two different control strategies in a smart home energy management system is demonstrated on the aging of a Lithium iron phosphate (LFP) battery.

## I. INTRODUCTION

The adoption of photovoltaic (PV) systems in the residential sector has been steadily increasing due to falling PV prices, financing incentives, programs targeting low-to-moderate income households, and a maturing PV market [1]. Even for the average-sized photovoltaic system, the mismatch between solar power production and simultaneous household loads could result in a significant power surplus, especially during peak daytime solar irradiance [2]. This mismatch can be addressed by integrating appropriately sized lithium-ion batteries (LIBs), so that this excess power generated can be stored and used in periods of insufficient solar power, instead of drawing power from the electricity grid. This has also become increasingly viable economically because of a continued trend of decreasing prices, with an 89% drop in the cost per kWh between 2010 and 2020 [3]. Energy storage (ES) systems paired with residential PV systems are even reported to compete with grid prices when appropriately sized and with government incentives [4].

Another approach to the problem of mismatched residential solar power generation and consumption is to schedule flexible loads to periods of solar power surplus. This is further encouraged by demand-response programs adopted by utility companies to flatten the load-curve of the grid, when accounting for reduced grid energy demand during day-time and the increased ramp-up rates in the late evening as solar irradiance decreases [5]. A Home Energy Management System (HEMS) can achieve load-levelling by scheduling deferrable loads such as laundry, dishwasher, and electric vehicle charging to periods of low grid load [6].

In a household equipped with PV and ES systems, the HEMS would additionally be capable of determining the charge-discharge strategy for the utilization of the ES system with regard to time of use pricing set by utility companies and free power available from the PV panels. With respect to control strategies for such residential ES systems, using the generated electricity instantly whenever possible is preferred instead of charging the battery, since charging, storing, and discharging energy are each associated with losses [7].

An important factor considered by potential ES adopters is the useable lifetime of the battery, which is associated with the return on investment. Battery aging with capacity fade is also crucial to an HEMS because it reduces the ability of the battery to meet the energy demands of the home appliances. Analysis of the aging phenomenon in LIBs reveals two nonlinear mechanisms, namely an increase in the internal resistance and a decrease of useful battery capacity [8].

In residential applications, control strategies do exist that aim to maximize battery lifetime but are largely focused on reducing battery dwell times near high states of charge (SOCs), which can cause premature aging [7]. A HEMS that considers the aging phenomenon dynamically using an aging model will be able to maximize the useful life more effectively, and for real-time application must be computationally efficient.

In this paper, a semi-empirical physics-based model is proposed that can easily be adapted to numerous battery chemistries and requires minimal computation effort, so that it can be run in real-time. Based on this approach, a comparison is conducted on the effect of energy management strategies on the ES battery life.

## II. LITERATURE REVIEW OF BATTERY AGING MODELS

When integrating the ES system into an HEMS, the battery will be periodically charged from either renewable or grid power sources, and then discharged to supply energy to the appliances. This cycling behavior ages the battery in addition to the calendar aging mode [9, 10]. The battery aging causes capacity decay and resistance increase which reduces the energy and power performance of a battery respectively [11].

There has been a lot of research working on understanding and mitigating battery aging recently, using both experimental and modeling techniques [12, 13]. In terms of modeling the battery aging phenomenon, different approaches have been utilized ranging from physics-based PDE models to data driven approaches [10, 12, 20]. A comparison of different lumped parameter models has been presented in [26].

*Research supported by The Ford Motor Company
C. Miller, M. Goutham, X. Chen, & S. Stockar Authors are with The Ohio State University, Mechanical & Aerospace Engineering Depart., Center for Automotive Research, Columbus, OH 43201 USA (C. Miller, University Fellow, provided phone: 419-835-4140, e-mail: miller.8184@osu.edu).
P.D. Hanumalagutti, R. Blaser Authors are with The Ford Motor Company, Dearborn, MI 48126 USA (R. Blaser provided e-mail: rblaser1@ford.com).

Physics-based battery aging models are built starting from the basic electrochemical mechanisms in a battery that are related to aging process. Such mechanisms include the increase of a solid electrolyte interphase (SEI) layer [10], the loss of active materials (LAM) [17], lithium plating [14], and cracks in the SEI layer [15,16]. Fortunately, the physics can be captured by modeling for most of the aging mechanisms such as SEI layer growth due to the side reaction between lithium and the electrolyte salt using Tafel and Nernst equations [10]. The crack in either the active material or the SEI layer can be predicted using equations describing the relations between stress, strain, displacement, or volume change [15]. Although more details included in the model equations will improve the accuracy of modeling, the computational complexity is increased, which is more problematic when applying these models for control applications such as in an HEMS. This is because the time scales of the electrochemical events inside the battery are much faster compared to dynamics of the energy management systems.

Fully data driven battery aging models differ significantly from physics-based models because they rely entirely on experimental or real-time feedback data. Simple data-driven models implement generic data fitting equations such as an empirical aging law [18], or the Arrhenius law equation [19]. These simple models follow rudimentary calibration techniques and result in easy implementation with a minimal number of calibration parameters. In contrast, more data-driven approaches have been developed where a deep learning algorithm is used to emulate the aging phenomenon [20], or where the remaining useful battery life is determined via a Monte Carlo simulation algorithm [21]. While data-driven aging model approaches can reduce computation time significantly, they always lack confidence in results outside of their calibration range.

In the third category, semi-empirical aging models are a mix of physics-based and data-driven approaches. These models simplify the physical process in those battery aging mechanisms with assumptions in either the limiting factors or the working conditions such as the C-rate, temperature, or SOC values [12, 16]. Most semi-empirical models further decouple battery aging from the original electrochemical reactions because of the distinct time scales for the two processes. With this, the aging models are simplified into feedforward models without affecting the battery model parameters over a short duration [12].

Comparing the three categories of battery aging models, this paper selects the semi-empirical model as the option considering the complexities in both computation and calibration with limited amount of data. The only limitation is the similarity required between the operating conditions used for calibration and actual application in the HEMS.

## III. MIXED-DEGRADATION MODEL

### A. Development of Mixed-Degradation Model

A mixed approach was chosen to model battery aging that utilizes the respective benefits from the data-driven calibration approach [18] and a semi-empirical physics-based model [12, 23]. The proposed Mixed-Degradation Model employs physics-based equations that model two primary battery aging mechanisms, SEI layer growth and LAM, that have also been utilized in numerous other semi-empirical models [12, 23]. The equations that model these mechanisms are (1,2) along with (3) which depicts the total capacity loss in the battery:

$$Q_{SEI} = \int_0^t \frac{k_{SEI} \exp\left(-\frac{E_{SEI}}{RT}\right)}{2(1+\lambda\theta)\sqrt{t}} dt \quad (1)$$

$$Q_{AM} = \int_0^t k_{AM} \exp\left(-\frac{E_{AM}}{RT}\right) \cdot SOC \cdot |I| dt \quad (2)$$

$$Q_{total} = Q_{SEI} + Q_{AM} \quad (3)$$

Except $\Theta$, which captures the effects of the reaction overpotential, all other parameters of the equations are available as constant battery characteristics, calibration parameters, or inputs from an equivalent circuit model. In the studied literature which implements these equations, the parameter $\Theta$ is deduced from complex electrochemical models in which partial differential equations must be solved, such as an extended single particle model [12, 23]. Therefore, computational efficiency is low in the context of real-time control applications. Additional context of the parameter $\Theta$ can be obtained in the references provided [12, 23].

The proposed model differs because, rather than solving for $\Theta$, it is instead chosen as an additional calibration parameter. Given that $\lambda$ itself, is also a calibration parameter, $\lambda$ and $\Theta$ are lumped as a single parameter given by $X := \lambda \cdot \Theta$. The new equation for capacity loss due to SEI layer growth is shown below as (4):

$$Q_{SEI} = \int_0^t \frac{k_{SEI} \exp\left(-\frac{E_{SEI}}{RT}\right)}{2(1+X)\sqrt{t}} dt \quad (4)$$

The procedure to calibrate the five model parameters, $k_{SEI}$, $E_{SEI}$, $X$, $k_{AM}$, and $E_{AM}$, to a specific battery type follows a procedure similar to that of [18] during calendar aging calibration. The proposed procedure is detailed below:

First, the parameters from (4), $k_{SEI}$, $E_{SEI}$, and $X$ are calibrated to experimental calendar aging data for the specific battery chemistry type by minimizing the deviation between model and experimental data. For the following steps, $k_{SEI}$ and $E_{SEI}$ are held constant at any other battery operating conditions.

Next, $X$ must be calibrated against additional calendar aging data sets in which the battery SOC and temperature are varied. This initial calendar aging procedure will yield a mapping of $X$ at different battery temperatures and SOCs allowing for easy interpolation and extrapolation. Typically, extrapolation is considered to be non-reliable, but because the mixed-degradation model is physics-based, there is additional confidence in extrapolating outside of the calibration zone.

The final step is to calibrate parameters from (2), $k_{AM}$ and $E_{AM}$ to experimental cycling aging data using a similar approach as the first step. Any obtained experimental cycling aging data is a combination of both calendar and cycling aging ($Q_{SEI}$ & $Q_{AM}$). Thus, during this calibration step, error should be minimized between experimental data and $Q_{total}$, given by (3).

The calendar aging calibration approach requires an assumption to be made that X, which is a function of SOC and temperature, is not dependent on C-rate. It is clear from [23], however, that the C-rate does play a role in determining Ө which is a function of reaction overpotential. Therefore, if the respective cycling current profile for cycling aging calibration differs from the validation current profile, then an error will be inherited.

The following sections implement the calibration procedure on a 2.3 Ah LFP Battery (26650) from A123s systems. This battery was also calibrated in [12, 17, 23].

### B. Calibration of Capacity Loss Due to SEI Layer Growth

In accordance with the first step in the procedure, the 2.3 Ah battery was calibrated to experimental calendar aging data due to SEI layer growth from [12, 17]. The calibrated parameters are shown below in Table 1.

TABLE 1. $Q_{SEI}$ Model Coefficients for LFP Battery Type

| Fitted Parameters | Value | Unit |
|---|---|---|
| $k_{SEI}$ | 7,350 | $1/\sec^{1/2}$ |
| $E_{SEI}$ | 39,333 | J/mol |

Per the second step of the procedure, the SEI layer growth coefficients were then held constant and used during the calibration of X at four additional operating conditions against calendar aging data also from [12, 17]. To increase the accuracy of the model over a larger operating range, (4) was calibrated at additional operating conditions in which two more experimental data sets were obtained from [23]. The results of the X calibration at each operating condition is shown below in Table 2.

TABLE 2. Tuned X at Various Operating Points for LFP Battery Type

| SOC (%) / T°C | 25°C | 30°C | 45°C | 60°C |
|---|---|---|---|---|
| **30%** | - | 1.6227 | 1.0331 | - |
| **50%** | 0.6970 | - | 0.2841 | - |
| **100%** | 0.0482 | - | 0.0331 | -0.1433 |

By obtaining these values of X at each operating condition, Table 2 can be implemented in the model as a lookup table. This allows for X to be interpolated & extrapolated at any operating condition during simulation. Fig. 1 demonstrates this mapping of X in a 3D plot.

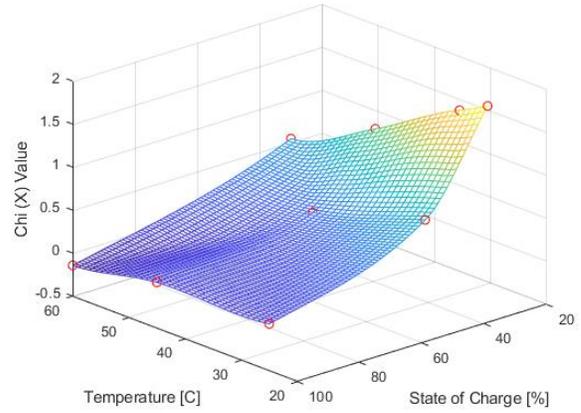

Figure 1. Mapping of Calibrated X Parameters at Different Battery Storage Conditions During Calendar Aging

To demonstrate the fitness of (4) after the calibration, Fig. 2 below displays the experimental and model's capacity loss due to SEI layer growth for five of the calibration sets.

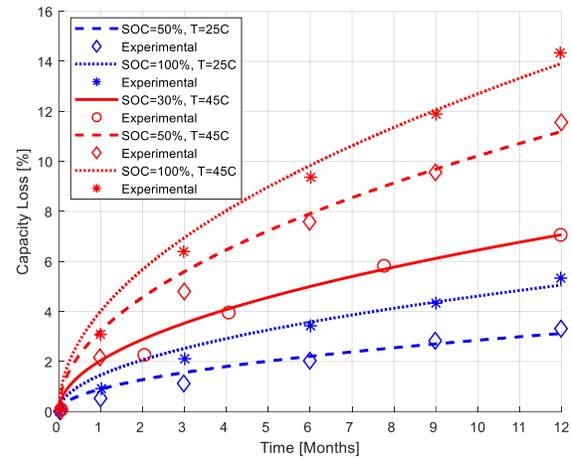

Figure 2: Calibration of SEI Layer Growth Equation during Calendar Aging Under Various Battery Storage Conditions Against Experimental Data Obtained from [12, 17]

### C. Calibration of Capacity Loss Due to LAM

Following the final step of the procedure, parameters in (2) must be calibrated for capacity loss due to LAM. To do so, experimental cycling aging data was obtained from [17]. This data represents total capacity loss (calendar & cycling) and therefore, $Q_{total}$ was fit against the experimental data for tuning of (2).

The experimental data for this calibration step comes from a study in which the battery is subject to a synthesized current profile which mimics a Hybrid Electric Vehicle (HEV) duty cycle [17]. This synthesized current profile was recreated using obtained data and implemented into our model shown below in Fig. 3. The profile subjects the battery to a wide range of C-rates, thus increasing the accuracy of the model as mentioned in the procedure. The results of the loss of active material calibration coefficients are shown below in Table 3:

TABLE 3. $Q_{AM}$ Model Coefficients for LFP Battery Type

| Fitted Parameters | Value | Unit |
|---|---|---|
| $k_{AM}$ | 1.1798 | 1/Ah |
| $E_{AM}$ | 39111 | J/mol |

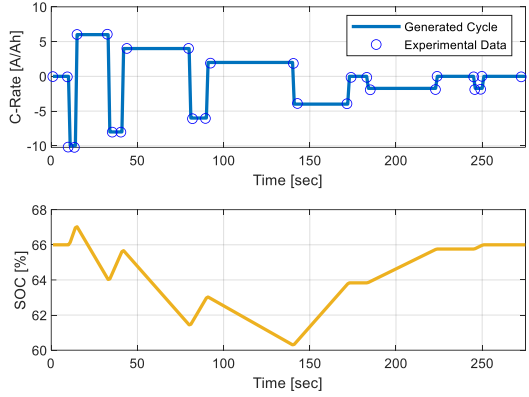

Figure 3: Synthesized current profile representing an HEV duty cycle obtained from [17]. Shown is one cycle which subject the battery to: Ah throughput per cycle = 0.48 Ah

To demonstrate the fitness of (4, 2) after the calibration, Fig. 4 below displays the experimental and model's total capacity loss.

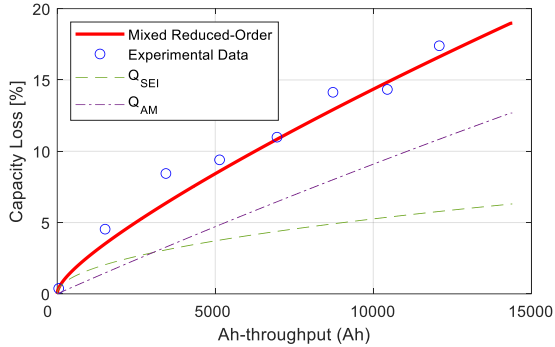

Figure 4: Simulation results of total capacity loss as a result of the calibrated parameters imposed on the mixed-degradation model compared to experimental results from [17]

### D. Validation of Calibrated Model

For the application of home energy management, current profiles of low C-Rate are typically imposed on the energy storage systems. Therefore, a data set resultant of a low C-Rate profile was used for model validation [25]. This imposed C-Rate cycling profile is shown in Fig. 5 which cycles the battery between 20% and 95%.

The results of the mixed-degradation model validation are shown in Fig. 6. Overall, the model tends to underestimate capacity loss and this deviation is attributed to the different profile and C-rate used for calibration as compared with that used for validation. Therefore, this error is expected, considering the major difference in calibration & validation duty cycles due to the scarcity of available data. It is important to note that for the utilization of this model, calibration of (2) should be done with an expected, application-specific current profile and respective experimental results.

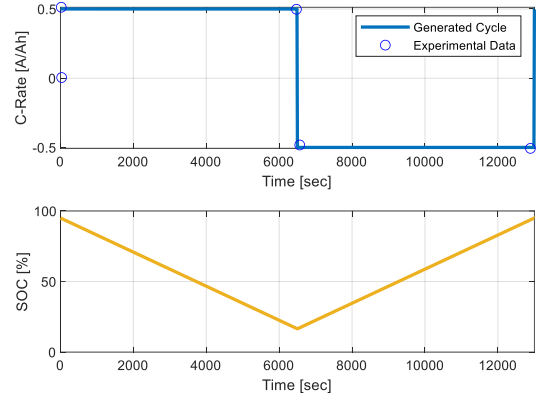

Figure 5: Synthesized current profile representing low C-Rate cycling obtained from [17]. Shown is one cycle which subject the battery to: Ah throughput per cycle = 3.6 Ah

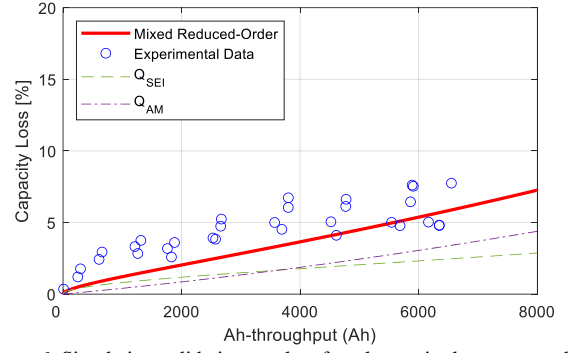

Figure 6: Simulation validation results of total capacity loss as a result of the calibrated parameters imposed on the mixed-degradation model compared to experimental results from [17]

### IV. CASE STUDY – HOME ENERGY MANAGEMENT SYSTEM

#### A. Background

The model used by [22] to simulate a Smart Home can vary a multitude of conditions to impose on the home such as location, home size (sq. ft.), number of household members and much more. The developed real-time energy management strategy can simulate the energy consumption of a smart house over a duration of time by taking into account the influence of the environment and activities of residents.

A baseline control policy also exists for the integration of the ES system with the smart home in order to store available surplus power generated by the photovoltaic system, thereby acting as an accumulator of energy. When the appliances in the household demand more energy than can be supplied by the photovoltaic system, the ES immediately discharges, while maintaining a minimum SOC of 20%.

The HEMS uses a more sophisticated optimization algorithm to schedule controllable and deferrable appliances in order to reduce electricity cost, while additionally determining the source of power to be used for these appliances, between the electricity grid and from the ES system. Considering this energy management strategy will utilize the SOC range of the ES more than the baseline policy, it is advantageous to integrate a battery aging model.

The home model previously discussed utilizes a first-order equivalent circuit model (ECM) to obtain the voltage and SOC of the ES battery for a certain current input and initial SOC. The SOC obtained by the ECM, the imposed current, and ambient battery temperature are used as inputs to the mixed-degradation model. This model was used to simulate long term use of the HEMS and extrapolate the capacity fade to predict battery end of life; ultimately, yielding the ability to size the battery for real world applications of such energy management systems. A summary of the integration is shown in Fig. 7.

The battery model does not currently include a thermal model to determine heat generated by the battery. This exclusion is based on the reasonable assumption that only low C-rates would be imposed on the battery based on the appliances in the household of the case study, yielding negligible heat generation. Therefore, the battery temperature is assumed to be the same as the home temperature if stored inside, or the ambient temperature if stored outside. However, future work or implementation within a different application could include a thermal model.

For this case study, the home was chosen to be located in Columbus, OH with a house size of 1606 sq. ft., a PEV battery size of 60 kW-hr, and energy storage battery size of 14 kW-hr.

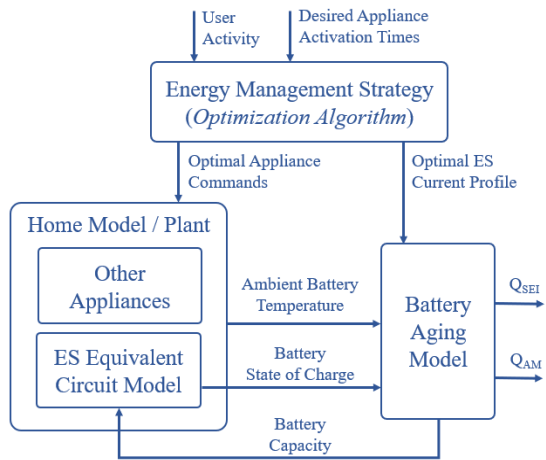

Figure 7: Home Energy Management System Integrated with Battery Aging Model Summary

*B. Case Study Results*

This paper analyzes how an HEMS affects the degradation of an integrated ES system over time when compared to a baseline policy for ES utilization (henceforth called "Baseline"). To gather meaningful results, a one-year simulation is conducted for the HEMS case and the Baseline. The metric applied is the end of life of the battery, defined as when the capacity depletes to 80% of nominal. To find when this occurs, the capacity fade profile obtained from the one-year simulation is extrapolated using a fitted function for the difference in capacity fade per year. Fig. 8 below shows the results for the first month capacity fade profile and Fig 9. shows the end-of-life estimation for the Baseline vs the HEMS.

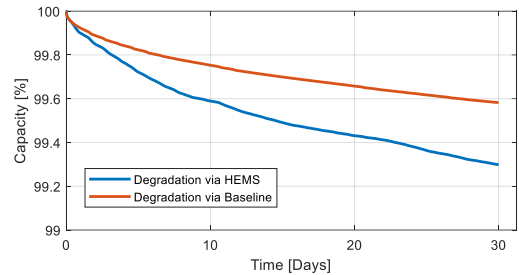

Figure 8: One month battery aging simulation results

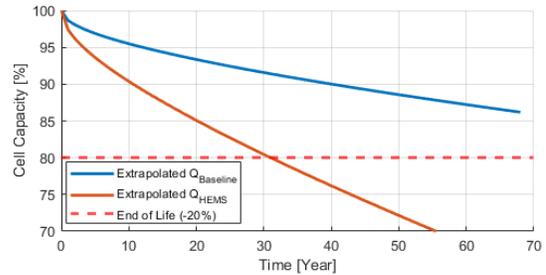

Figure 9: Battery End of Life Estimation

The HEMS end of life was estimated to be approximately 31 years while the Baseline was estimated to be approximately 69 years. In order to analyze the causes of these results in more detail, Fig. 10 and Fig. 11 show a histogram of the battery charge / discharge C-Rate and SOC probability respectively over the entire year. These figures yield insight into the demands being imposed on the battery in the HEMS case versus the Baseline case.

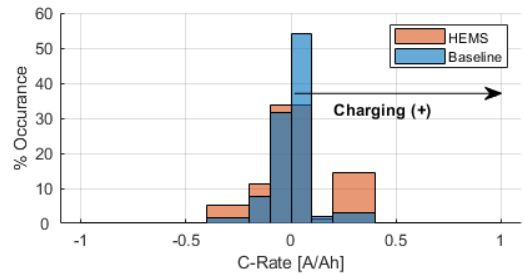

Figure 10: Histogram of Imposed Battery C-Rates

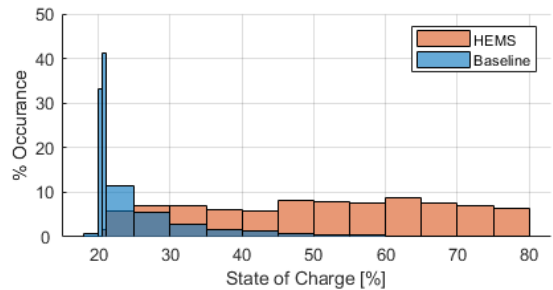

Figure 11: Histogram of Imposed Battery SOCs

Given that the aforementioned energy management strategy works to minimize grid cost, it will inherently impose more work on the battery; i.e., holding the battery at higher states of charge and demanding quicker discharge during high power requests. This effect is seen in Fig. 10 as the HEMS imposes much more frequent discharges and charges at higher C-Rates than in the Baseline case. This will inherently cause larger capacity loss due to loss of active material. Additionally, as seen in Fig. 11, the HEMS tends to sustain a higher SOC on the battery than the Baseline which will cause more capacity loss due to SEI layer growth and loss of active material.

## V. CONCLUSION

In this paper, a Mixed-Battery Degradation model was developed which combines the respective benefits of an entirely physics-based model and an empirical model. The model presented in this paper allows for easy calibration for different battery chemistry types and is computationally efficient enough such that it can be implemented in real-time control applications. In this paper, the degradation model is used in post-processing to evaluate and compare the effect of different home energy management strategy on capacity fade. As expected, as smart home algorithm will use the battery more aggressively and hence results in accelerated aging, compared to a rule-base strategy.

While the comparison between strategies provides helpful insight on battery usage, the predicted end of life is affected by the limitations of the model. Namely the calibration performed on data collected at higher C-rate; and the longer simulation times compared to automotive applications. The first results in model extrapolation, the latter in accumulation of error. Future work will focus on obtaining aging data that are specific for stationary applications.

ACKNOWLEDGMENT

The authors thank The Ford Motor Company for supporting this work.